\newcommand{\ale}{\ \raisebox{-.3ex}{$\stackrel{<}{\scriptstyle \sim}$}\ }
\newcommand{\age}{\ \raisebox{-.3ex}{$\stackrel{>}{\scriptstyle \sim}$}\ }

\documentclass{mn2e}

\input psfig

\title[Orbital radii of extrasolar planets]
	{Predictions for the frequency and orbital radii of massive extrasolar planets}
	
\author[P.J. Armitage, M. Livio, S.H. Lubow \& J.E. Pringle]{Philip J. Armitage$^{1,2}$,  
        Mario Livio$^3$, S.H. Lubow$^3$ and J.E. Pringle$^{4,3}$ \\
	$^1$JILA, University of Colorado, 440 UCB, Boulder CO 80309-0440, USA \\
	$^2$School of Physics and Astronomy, University 
	of St Andrews, North Haugh, St Andrews KY16 9SS, UK \\
	$^3$Space Telescope Science Institute, 3700 San Martin Drive, 
	Baltimore MD 21218, USA \\	
	$^4$Institute of Astronomy, 
	Madingley Road, Cambridge CB3 0HA, UK}	

\begin{document}

\maketitle

\begin{abstract}
We investigate the migration of massive extrasolar planets due to 
gravitational interaction with a viscous protoplanetary disc. 
We show that a model in which planets form at 
5~AU at a constant rate, before migrating, leads to a predicted 
distribution of planets that is a steeply rising function of 
$\log (a)$, where $a$ is the orbital radius. Between 1~AU and 
3~AU, the expected number of planets per logarithmic 
interval in $a$ roughly doubles. We demonstrate that, once selection effects are 
accounted for, this is consistent with current data, and   
then extrapolate the observed planet 
fraction to masses and radii that are inaccessible to 
current observations. In total, about 15\% of stars targeted 
by existing radial velocity searches are predicted to possess 
planets with masses $0.3 M_{\rm J} < M_{p} \sin (i) < 10 M_{\rm J}$ , and radii 
$0.1 {\rm AU} < a < 5 {\rm AU}$. A third of 
these planets (around 5\% of the target stars) lie at the radii most amenable 
to detection via microlensing. A further 5-10\% of stars could have  
planets at radii of $5 {\rm AU} < a < 8 {\rm AU}$ that have migrated outwards. 
We discuss the probability 
of forming a system (akin to the Solar System) in which 
significant radial migration of the most massive planet 
does {\em not} occur. About 10-15\% of systems with a 
surviving massive planet are estimated to fall into 
this class. Finally, we note that a smaller fraction of 
low mass planets than high mass planets is expected to 
survive without being consumed by the star. The initial 
mass function for planets is thus predicted to rise more 
steeply towards small masses than the observed mass function.
\end{abstract}

\begin{keywords}	
	accretion, accretion discs --- 
	solar system: formation ---
	planetary systems: formation --- 
	planetary systems: protoplanetary discs --- gravitational lensing
\end{keywords}

\section{Introduction}
Radial velocity surveys of nearby stars show that a significant 
fraction -- at least 8\% -- have massive planets with orbital 
radii substantially less than that of Jupiter (Marcy \& Butler 2000; 
Udry et al. 2000; Butler et al. 2001). Most of 
these planets lie at radii where massive 
planet formation is theoretically believed to be difficult (Bodenheimer, 
Hubickyj \& Lissauer 2000). The difficulties are most 
pronounced for those planets at the smallest orbital 
radii, where the temperatures in the protoplanetary 
disc would have exceeded those for which ices (and possibly 
even dust) can exist. These problems can be avoided 
if planets formed at larger radii, and then migrated 
inwards to where they are detected today. Mechanisms 
that could lead to this migration include gravitational 
interaction with a gaseous and viscous protoplanetary disc 
(Lin, Bodenheimer \& Richardson 1996), planet-planet 
scattering (Rasio \& Ford 1996; Weidenschilling \& Marzari 1996; 
Lin \& Ida 1997; Ford, Havlickova \& Rasio 2001; 
Papaloizou \& Terquem 2001), or scattering 
of planetesimals by massive planets (Murray et al. 1998).  

In the right circumstances, all of the suggested mechanisms 
can lead to substantial migration. More quantitative 
comparisons with the data are therefore warranted. Recently, 
Trilling, Lunine \& Benz (2002) have reported 
promising results for the protoplanetary disc migration model. 
They showed that the broad distribution of orbital radii 
was consistent with the planets becoming stranded, during 
their inward migration, by the dispersal of the protoplanetary 
disc. They used the model to estimate both the fraction of 
stars that must have formed planets, and how many ought 
to have survived migration.

In this paper, we extend the work of Trilling et al. (2002). 
We include a physical mechanism for disc dispersal into a 
model for the formation and migration of massive 
planets, and make quantitative comparison with the 
observed distribution of planetary orbital radii.
Our disc model, described in Section~2, combines viscous 
evolution with mass loss at the outer edge, for example 
due to photoevaporation (Johnstone, Hollenbach \& Bally 1998; 
Clarke, Gendrin \& Sotomayor 2001). We then run this model 
repeatedly, on each occasion adding a single planet to the 
disc at a specified formation time $t_{\rm form}$. By 
varying $t_{\rm form}$, subject to the constraint that 
no planets can be formed once the disc mass is too low, 
we study in Section~3 how the final orbital radii of planets depend 
upon the formation time. With the addition of plausible 
assumptions about how the rate of planet formation varies 
with time, we then convert the results into a prediction 
for the radial distribution of planetary orbits. In Section~4 we
compare this distribution with the current data, and find that good  
agreement is obtained. In Section~5 we make use of the model 
to estimate the fraction of stars targeted by radial velocity 
surveys that ought to harbour planets, including those with 
masses too low, or orbital radii too large, to be currently 
detected. We also predict the number of stars with planets 
at radii suitable for detection via microlensing, and the 
fraction of systems where significant planetary migration 
does not occur. Our conclusions are summarized in Section 6.

\section{Methods}

\subsection{Fully viscous disc model}

The simplest protoplanetary disc model that meets the basic observational 
constraints on disc mass, disc lifetime, and disc destruction 
time-scale is a viscous disc with mass loss from the outer regions 
(Hollenbach, Yorke \& Johnstone 2000; Clarke, 
Gendrin \& Sotomayor 2001). We use this model for the majority 
of the calculations described in this paper, since it 
has a minimal number of free parameters. 

For the viscous disc model runs, we assume 
that over the relevant range of radii (roughly, 
between 0.1~AU and 10~AU), the kinematic viscosity $\nu$ 
can be described as a power law in radius $R$,
\begin{equation}
 \nu = \nu_0 \left({ R \over {\rm 1 \ {\rm AU}} } \right)^\beta.
\label{eqnu} 
\end{equation} 
The initial conditions for the surface density are a 
constant accretion rate disc, with a profile which 
corresponds to the assumed viscosity of equation (\ref{eqnu}). 
At radii $R \gg R_{\rm in}$, the inner edge of the disc, 
this implies,
\begin{equation} 
 \Sigma = { \dot{M} \over {3 \pi \nu} } = \Sigma_0 R^{-\beta}
\end{equation}
where $\dot{M}$ is the accretion rate and $\Sigma_0$ a constant. 
For our standard disc 
model, we take $\beta = 3/2$ (Weidenschilling 1977; Hayashi 1981), 
and choose $\nu_0 = 1.75 \times 10^{13} \ 
{\rm cm}^{2} {\rm s}^{-1}$ to give a sensible time-scale for disc 
evolution of a few Myr. Finally, $\Sigma_0$ is set in the initial 
conditions so that the initial disc mass is $0.1 M_\odot$. 
The inner and outer boundaries for the calculation are 
$R_{\rm in} = 0.066 \ {\rm AU}$, and
$R_{\rm out} = 33.3 \ {\rm AU}$. To extend the predictions 
to smaller radii, a limited number of runs 
have also been completed with an inner boundary at 
$R_{\rm in} = 0.03 \ {\rm AU}$.
A zero torque boundary condition (i.e. a surface density 
$\Sigma = 0$) is applied at the inner boundary, 
while at the outer boundary we set the radial velocity 
to zero.

Current observations do not directly constrain the surface 
density profile of protoplanetary discs at the AU scale. 
To check how sensitive our results are to changes in this 
profile, we have also run a model with $\beta = 1/2$, 
again normalized to give a sensible time-scale for disc evolution. 
A flatter profile than the $R^{-3/2}$ of our standard model 
is expected if the angular momentum transport within the 
disc follows the Shakura \& Sunyaev (1973) $\alpha$-prescription, 
with a constant $\alpha$ (e.g. Bell et al. 1997).

Even a small mass of gas, trapped exterior to the orbit 
of a massive planet, will eventually soak up a large amount 
of angular momentum and drive migration. This gas 
has to be lost in order to obtain a converged final radius 
for migrating planets. 
Trilling et al. (2002) make the simplest assumption, and 
halt migration by removing the disc 
instantaneously. We adopt a different approach, and assume  
that the gas is removed by photoevaporation. In 
this process, the surfaces of the disc are heated by the 
absorption of ultraviolet radiation, which may originate either from the 
central star or from an external source (e.g. a massive 
star in a young cluster). The hot gas can escape as a wind 
at radii $R > R_g$, where the escape velocity is less than the 
sound speed of the heated material. Observations 
of the Orion nebula show that this process drives mass
loss at significant rates, at least in clusters containing 
massive stars (Bally, O`Dell \& McCaughrean 2000).

Our implementation of photoevaporation within a time-dependent 
disc model is based upon the description given by Johnstone, 
Hollenbach \& Bally (1998; see also Shu, Johnstone \& 
Hollenbach 1993; Richling \& Yorke 2000). The standard 
model has $R_g = 5 \ {\rm AU}$, and a radial distribution 
of mass loss,
\begin{eqnarray} 
 \dot{\Sigma} &\propto& R^{-1} \, \, \, \, \, R > R_g \nonumber \\
 \dot{\Sigma} &=& 0 \, \, \, \, \, R < R_g. 
\end{eqnarray}
For a disc with an outer radius $R \gg R_g$, this means that 
the total mass loss scales linearly with $R$.
The initial mass loss rate (integrated over the disc out to 25~AU) 
is chosen to be $\dot{M} = 5 \times 10^{-9} M_\odot {\rm yr}^{-1}$.

\begin{figure}
\psfig{figure=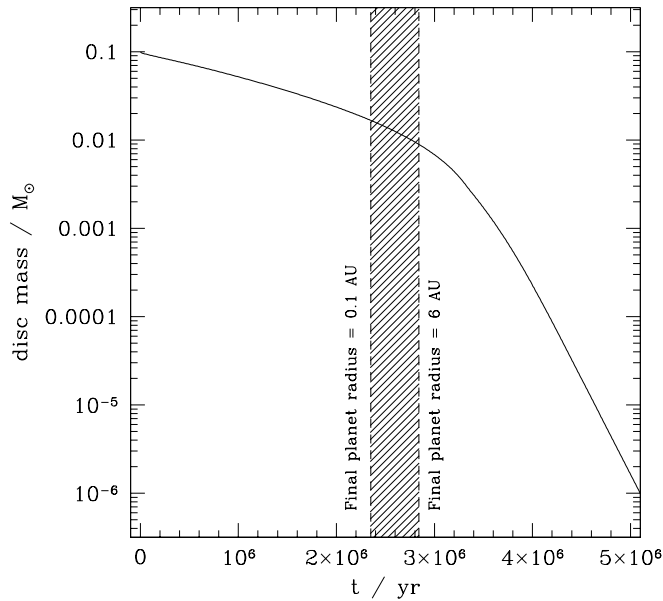,width=3.5truein,height=3.5truein}
\caption{Evolution of the mass of the standard disc 
	model with time. Up to around 3 Myr, the disc 
	mass decays as gas is accreted onto the star, 
	while the effects of the disc wind remain small.
	At later times, a steeper decline occurs 
	as mass loss truncates the outer radii of the 
	disc. The shaded band shows the range of time 
	over which a $2 M_J$ planet can be formed at 5~AU 
	and survive with a final orbital semi-major axis 
	$0.1 {\rm AU} < a < 6 {\rm AU}$.} 
\label{fig1}	
\end{figure}

Figure \ref{fig1} shows how the disc mass changes with time, for 
the standard model with $\beta = 3/2$ and mass loss from 
$R > 5 {\rm AU}$. The disc mass declines from 
$0.1 M_\odot$ to $10^{-2} M_\odot$ in 3~Myr, mainly 
due to accretion, before 
dropping rapidly once the effects of the disc wind 
take hold. Although the power laws and location of the turnover 
vary with the adopted parameters, generically similar behavior 
occurs for a range of disc viscosity and mass loss rates 
(Clarke, Gendrin \& Sotomayor 2001).

\subsection{Magnetically layered disc model}

In order to check the robustness of the results, we investigate how
the results would change if we adopted a fundamentally different,
but strongly theoretically motivated, model for disc evolution.
The magnetically layered model proposed 
by Gammie (1996) is based upon the observation 
that cold gas at $T \ale 10^3 {\rm K}$ has a low 
ionization fraction, which suppresses angular momentum 
transport via magnetohydrodynamic  
turbulence (Fleming, Stone \& Hawley 2000). If there 
are no other sources of angular momentum transport, then only 
the hot inner disc, plus a surface layer ionized 
by cosmic rays, will be viscous. At radii $R \sim 1 {\rm AU}$, 
this implies that accretion occurs only through a thin 
surface layer, below which lies a thick `dead zone' of 
quiescent gas. Numerical models for the long-term 
disc evolution (Armitage, Livio \& Pringle 2001) 
find that there is an early phase, lasting for 
perhaps a Myr, during which accretion 
occurs in short high accretion rate outbursts. This 
is followed by a quiescent phase, characterised by low rates 
of accretion onto the star. This is due to a much 
reduced (compared to fully viscous disc models) viscosity 
at radii of the order of 1~AU.

Detailed models for the layered disc are able to provide 
a reasonable fit to observations of the accretion rate 
in Classical T Tauri stars (Gammie 1996; but see also 
Stepinski 1998). For the purpose of studying planetary 
migration, we adopt an approximate approach, and assume 
that the most important effect of the layered disc is 
to reduce the viscosity near the midplane. Specifically, 
after 1~Myr, we reduce the viscosity where the surface 
density exceeds the thickness of the magnetically active 
surface layer. The new viscosity $\nu_{\rm layer}$ is,
\begin{equation}
 \nu_{\rm layer} = \nu \times {\rm min} \left( { {100 {\rm gcm}^{-2}} \over \Sigma} , 1 
 \right)
\end{equation}
where $\nu$ is as given in equation (\ref{eqnu}), and the same viscosity 
parameters as for the standard disc model are chosen. To obtain a 
sensible disc lifetime we also  
increase the mass loss rate to $\dot{M} = 3 \times 10^{-8} M_\odot {\rm yr}^{-1}$, 
and allow mass to be lost at all radii by taking $R_g = R_{\rm in}$. 

\subsection{Planet formation assumptions}

The location in the disc where massive planet 
formation is easiest depends upon  
two competing factors. The characteristic time-scales 
for planet building in the disc are faster at small radii.
However, most of the mass in protoplanetary discs lies at 
large radii, with a jump in the surface density of 
solid material beyond the `snow line' where ices 
can form (Hayashi 1981; Sasselov \& Lecar 2000). We 
assume that the outcome of this competition is that 
planets form at an orbital radius $a = 5 {\rm AU}$, 
similar to that of Jupiter in the Solar System.
The exact choice of this location is less important 
than the fact that it lies outside the radii where 
extrasolar planets are currently observed. In the 
model which we are testing, this means that 
migration is necessary to explain the orbital radii 
of all currently known planets.
 
In each run of the disc model, we form a planet with 
mass $M_p$, at time $t_{\rm form}$ (where $t=0$ is defined 
as the time when the disc mass is $0.1 M_\odot$). We 
specify the mass of our planets (typically $2 M_J$, 
where $M_J$ is the mass of Jupiter) in advance of 
each run, but test at the time of formation 
that there is enough mass in the vicinity of the 
planet to form it consistently. Specifically, we 
check that the disc mass between $R=0.6a$ and $R=1.6a$ 
exceeds the desired 
planet mass. Provided that this is satisfied, 
we remove the appropriate amount of gas from 
the disc around $R = a$, and add a planet with 
mass $M_p$ at that location. This 
instantaneous formation scheme ignores the 
possibly lengthy time required to assemble the 
planet's gaseous envelope (e.g. Lissauer 1993, 
and references therein). However, Trilling, Lunine \& 
Benz (2002) have shown that including a gradual 
build-up of mass makes little difference to the 
outcome of migration.

After running the model a number of times, the basic 
output is a plot of the final radius of a planet as 
a function of the time when it formed. To convert this 
to the number of planets expected at different radii, 
which is the observable quantity, we also need to specify 
how the probability of massive planet formation, per unit 
time, varies with time. We assume that this probability 
is uniform (or equivalently, that the rate of massive planet 
formation, averaged over many stars, is constant). 
This cannot be true over long periods of 
time. However, we are only interested in the 
window of formation times which allow a massive 
planet to survive migration. This window, shown 
as the shaded region in Figure \ref{fig1}, is 
short compared to the disc lifetime, and as a result there is 
only a relatively small change in the physical 
properties of the disc during this time. For 
example the disc mass, 
which drops by two orders of magnitude over 
the first 3.5~Myr, changes by less than 50\%.  
Whatever factors influence the probability of planet 
formation, it should therefore be a reasonable 
first approximation to assume that the probability 
is constant over the short range of formation times 
which lead to a surviving massive planet.

\subsection{Model for planetary migration}

Our model for planetary migration is almost 
identical to that of Trilling et al. (2002), 
who describe in detail the approximations involved. 
The coupled evolution of the planet and the 
protoplanetary disc is described by the 
equation (Lin \& Papaloizou 1986), 
\begin{equation} 
 { {\partial \Sigma} \over {\partial t} } = 
 { 1 \over R } { \partial \over {\partial R} } 
 \left[ 3 R^{1/2} { \partial \over {\partial R} } 
 \left( \nu \Sigma R^{1/2} \right) - 
 { { 2 \Lambda \Sigma R^{3/2} } \over 
 { (G M_*)^{1/2} } } \right].
\label{eqsigma} 
\end{equation}   
The first term on the right-hand side of this equation 
describes the diffusive evolution of the surface 
density due to internal viscous torques 
(e.g. Pringle 1981). The second term describes how 
the disc responds to the planetary torque, $\Lambda (R, a)$, 
where this function is the rate of angular momentum 
transfer per unit mass from the planet to the disc.
We take, 
\begin{eqnarray} 
 \Lambda & = & - { {q^2 G M_*} \over {2 R} } 
 \left( {R \over \Delta_p} \right)^4 \, \, \, \, R < a \nonumber \\
 \Lambda & = & { {q^2 G M_*} \over {2 R} } 
 \left( {a \over \Delta_p} \right)^4 \, \, \, \, R > a 
\label{eqtorque} 
\end{eqnarray} 
where $q = M_p / M_*$, 
\begin{equation}
 \Delta_p = {\rm max} ( H, \vert R - a \vert ),
\end{equation} 
and $H$ is the scale height of the disc. Guided by 
detailed protoplanetary disc models (Bell et al. 1997), 
we adopt $H = 0.05 R$. This form for $\Lambda$ (equation 
\ref{eqtorque}) is that used by Lin \& 
Papaloizou (1986), modified to give a symmetric 
treatment for the disc inside and outside the 
orbit of the planet. 

The transfer of angular momentum leads to orbital 
migration of the planet at a rate,
\begin{equation}
 { { {\rm d} a } \over { {\rm d} t } } = 
 - \left( { a \over {GM_*} } \right)^{1/2} 
 \left( { {4 \pi} \over M_p } \right) 
 \int_{R_{\rm in}}^{R_{\rm out}} 
 R \Lambda \Sigma {\rm d} R.
\end{equation} 
Although simple, this 
formalism for treating the planet-disc interaction 
has been shown (Trilling et al. 1998) to give 
comparable results to more sophisticated methods 
(Takeuchi, Miyama \& Lin 1996).

We solve equation (\ref{eqsigma}) using an 
explicit method on a grid that 
is uniform in a scaled variable $\propto R^{1/2}$, with 
300 grid points between $R_{\rm in}$ and $R_{\rm out}$. 
The stellar mass is $M_* = M_\odot$.
Typically, the timestep is limited by 
the radial velocity in the disc gas very near the location of 
the planet. Evolving the system on this (small) timestep 
serves no useful purpose once a gap has been opened, so 
we reduce $\Lambda$, and thereby limit $\vert v_r \vert$, in the 
vicinity of the planet. The results of the migration 
calculations are unaffected by this modification.

\section{Results}

\subsection{Predicted semi-major axis of stranded planets}

\begin{figure}
\psfig{figure=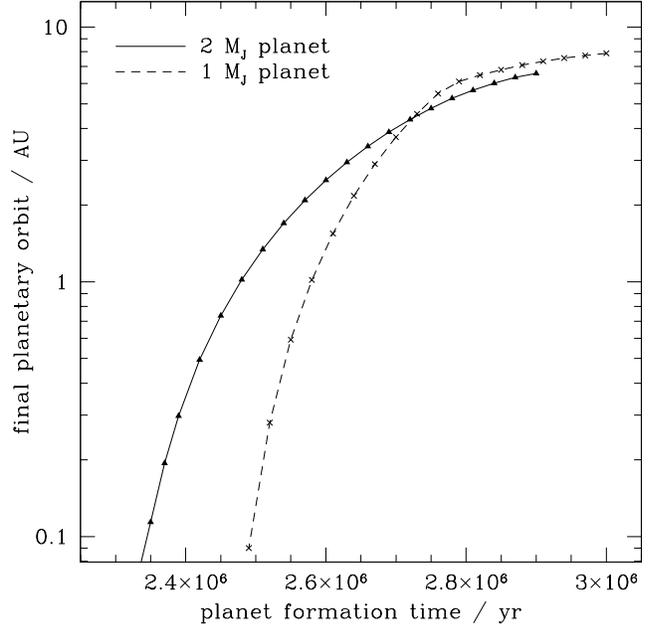,width=3.5truein,height=3.5truein}
\caption{The final semi-major axis of planets, following 
	migration, as a function of the planet formation 
	time (i.e. the epoch when the planet formed at 
	5~AU). The standard disc model with $\beta = 3/2$ 
	and mass loss at $R > 5 {\rm AU}$ was used. The solid 
	curve shows results for $2 M_J$ planets, the dashed 
	curve results for $1 M_J$ planets. More massive 
	planets can form earlier without being consumed 
	by the star.} 
\label{fig2}	
\end{figure}

Figure \ref{fig2} shows the final semi-major axis of planets as 
a function of the formation time, using the standard disc 
model of Fig.~\ref{fig1}. Each point  
represents a separate run of the model. The fate of a planet depends 
upon when it was formed, and upon its mass. Planets 
that are formed too early migrate inward of 0.1~AU, 
the smallest radius we consider quantitatively in this study. What 
happens to them subsequently depends upon whether there 
is a stopping mechanism to halt further migration, but 
many of them seem likely to be consumed by the star. 
Planets that start to form too late can migrate 
outwards, because at late times the gas in the 
disc at 5~AU is itself moving away from the star.
At still later times, it becomes impossible to form a 
Jupiter mass planet at all, because the dispersing disc has too 
little surface density (c.f. the discussion in Shu, Johnstone \& 
Hollenbach 1993). In between these limits, there is a 
window of formation times, lasting for around 0.5~Myr 
for a $2M_J$ planet, which result in the planet being 
stranded at radii between 0.1~AU and 6~AU. The migration 
of more massive planets slows down earlier, since their 
larger angular momentum produces greater resistance to 
disc-induced migration (Syer \& Clarke 1995). Hence, the 
window of allowable formation times is longer for 
more massive planets than for less massive ones. 

\subsection{Dependence upon disc model and planet mass}

\begin{figure}
\psfig{figure=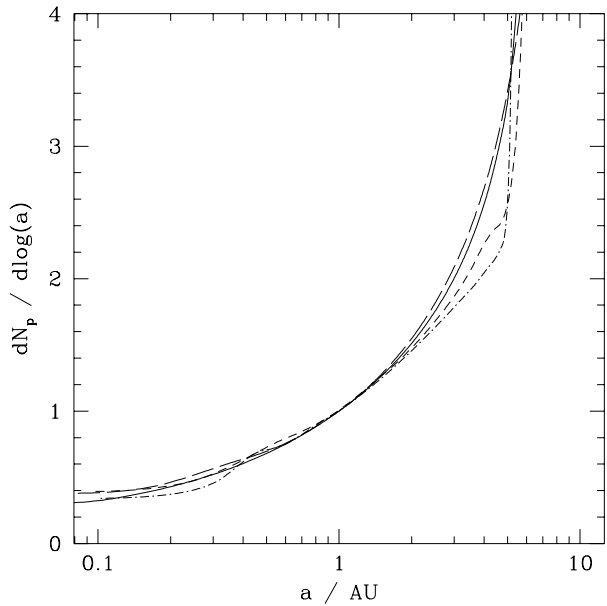,width=3.5truein,height=3.5truein}
\caption{Predicted number of extrasolar planets per logarithmic 
	interval in semi-major axis. These curves are derived 
	from the $a(t_{\rm form})$ results shown in Fig.~3 by 
	making the additional assumption that the rate of 
	planet formation is constant. The solid curve shows results 
	for the standard disc model ($2 M_J$ planets, $\beta=3/2$, 
	$R_g = 5 {\rm AU}$). The remaining curves show the effect of changes 
	to these parameters: a lower planet mass $M_p = 1 M_J$ (short 
	dashes), increased radius of mass loss $R_g = 10 {\rm AU}$ 
	(long dashes), and different viscosity $\beta=1/2$  (dot-dashed).
	Because the $\beta=1/2$ model has a lower surface density at 
	5~AU, the planet mass used was reduced to $0.5 M_J$.
	The absolute number of planets is arbitrary, and has been 
	normalised to unity at 1~AU.}
\label{fig3}	 
\end{figure}

Figure \ref{fig2} shows how the 
final planetary radius depends upon the formation time (note that we plot 
$a(t_{\rm form})$ on a logarithmic scale in this figure). The curve 
steepens towards small radii / early formation times.
This implies that if planet formation occurs with 
uniform probability per unit time during the allowed 
window, then fewer planets per logarithmic interval in $a$ 
are expected at small radius (i.e. equal intervals 
in $\log(a)$ correspond to smaller intervals in 
$\Delta t_{\rm form}$ at small $a$). 

With the assumption of a uniform rate of planet formation, 
the predicted number of planets per logarithmic interval 
in radius is,
\begin{equation}
 { { {\rm d}N_p } \over { {\rm d} \log (a) } } \propto
 \left( { { {\rm d} \log (a) } \over { {\rm d} t_{\rm form} } } 
 \right)^{-1}, 
\label{eqdnda} 
\end{equation}
which can readily be derived from Fig.~\ref{fig2}. Figure \ref{fig3} shows the 
results for two different planet masses, and for two 
variants on the standard disc model. A cubic spline fit 
has been used to obtain a smooth estimate of the derivative 
in equation (\ref{eqdnda}), and the curves have been normalised to 
unity at 1~AU. 

The predicted 
number of planets per logarithmic interval in $a$ 
rises rapidly with increasing radius for all the models considered. For the 
standard disc model, the 
predicted number rises by more than a factor of 
two between 0.1~AU and 1~AU, and by a further factor 
of two by 3~AU. 
The good agreement between the curves for $1 M_J$ and 
$2 M_J$ planets shows that the {\em observed} mass 
function for massive planets is predicted to be 
the same at different radii. This means that we can 
use the observed mass function at small radii, which 
is complete down to lower masses, to estimate how 
many low mass planets must accompany known high 
mass planets at larger radii. It does not mean that 
the initial mass function for planets is the same 
as the observed one. In fact, Fig.~\ref{fig2} makes it 
clear that high mass planets can form over a 
longer interval than low mass planets and still 
survive migration. To end up with the same 
number of $1 M_J$ as $2 M_J$ planets per unit 
interval in $\log(a)$, more low mass planets must 
have formed. In other words, the (currently unobservable) 
initial mass function for planets must rise more 
steeply towards low masses than the observed one.

Figure~\ref{fig3} also shows the predicted relative abundance of 
planets for two variants of the disc model. Increasing the 
inner radius beyond which mass is lost to 10~AU does not alter 
the predicted distribution of planets within 5~AU. Likewise, the 
predicted distribution is not substantially changed in a model 
where the power-law slope of the viscosity is $\beta=1/2$. Note 
that for the $\beta=1/2$ run we considered planets with mass 
$0.5 \ M_J$, since insufficient gas was available between 
$0.6 a$ and $1.6 a$ to form planets of several Jupiter masses. 

\subsection{Comparison with an instantaneous disc dispersal model}

\begin{figure}
\psfig{figure=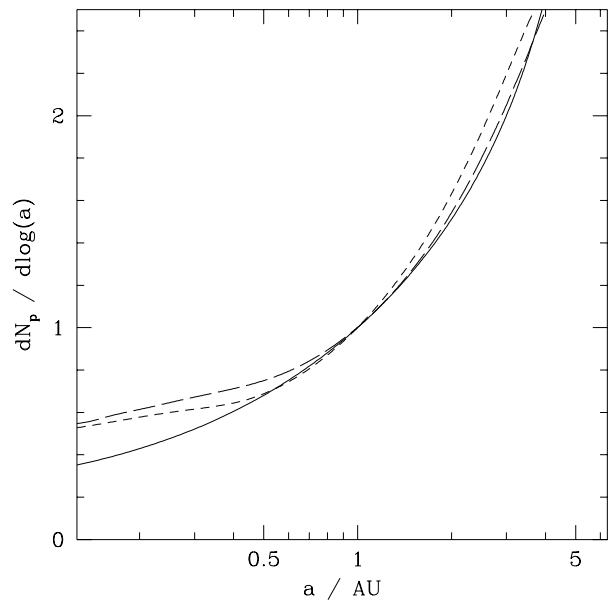,width=3.5truein,height=3.5truein}
\caption{Predicted number of extrasolar planets per logarithmic 
	interval in semi-major axis. The solid line shows 
	the results for the standard disc model, as in 
	Fig.~3. The short-dashed and long-dashed curves show 
	models in which the disc is 
	lost instantaneously at $t = 3~{\rm Myr}$, or 
	$t = 4~{\rm Myr}$, respectively. The predicted distribution  
	of planets that have migrated {\em inwards} is almost 
	the same in all of these models.}
\label{fig_instant}	 
\end{figure}

In Figure \ref{fig_instant}, we show how the model including disc mass 
loss compares to a simpler model in which the disc is assumed to be lost 
instantaneously at a time $t_{\rm disperse}$. The disc model used is identical 
to our standard model with $\beta = 3/2$, except that the rate 
of mass loss is set to zero. We compute two models, with $t_{\rm disperse}$  
equal to 3~Myr and 4~Myr respectively. 

From Fig.~\ref{fig_instant}, it can be seen that the instantaneous 
disc dispersal model makes very similar predictions for the radial 
distribution of inwardly migrating planets. Larger differences, 
however, occur for outward migration. In the instantaneous disc 
dispersal model, there is no outward migration of planets, because 
the radial velocity in the disc at 5~AU remains inward at 
$t_{\rm disperse}$. The model including mass loss, conversely, 
allows a substantial fraction of planets to migrate outwards. 
From Fig.~\ref{fig2}, we find that for Jupiter mass planets 
as many as a third may end up in orbits exterior to that 
in which they formed. 

\subsection{Effect of ongoing planetary accretion}

Even after a gap has been opened, numerical simulations show 
that planetary accretion may continue (Bryden at al. 1999; 
Kley 1999; Lubow, Seibert, \& Artymowicz 1999; Nelson et al. 2000). 
Significant ongoing 
accretion -- which is not included in our models -- would 
lead to a correlation between planet mass and orbital 
radius, with more massive planets expected at small 
orbital radius. There is no evidence for such a 
correlation in the data, which could either indicate 
that migration has not occurred, or that ongoing 
accretion occurred only at a low level.

Lubow, Seibert, \& Artymowicz (1999) find that, for low 
masses, the rate of ongoing accretion can be comparable 
to the {\em disc} accretion rate outside the planet.
This is of the order of $10^{-9} \ M_\odot {\rm yr}^{-1}$ 
at the epoch when surviving planets form. Migration from 
5~AU takes a few hundred thousand years, so we estimate 
that accretion across the gap could add a few tenths of 
a Jupiter mass to low mass planets. This is a relatively 
small fraction of the mass for the $2 M_J$ planets that 
are the main focus of this paper. However, it is only 
marginally consistent with the existence of the lowest 
mass hot Jupiters, which have $M_p \sin (i) < 0.3 \ M_J$.
This may indicate that these planets formed within the 
snow line (leading to a shorter migration time).
 
\subsection{Magnetically layered protoplanetary discs}

\begin{figure}
\psfig{figure=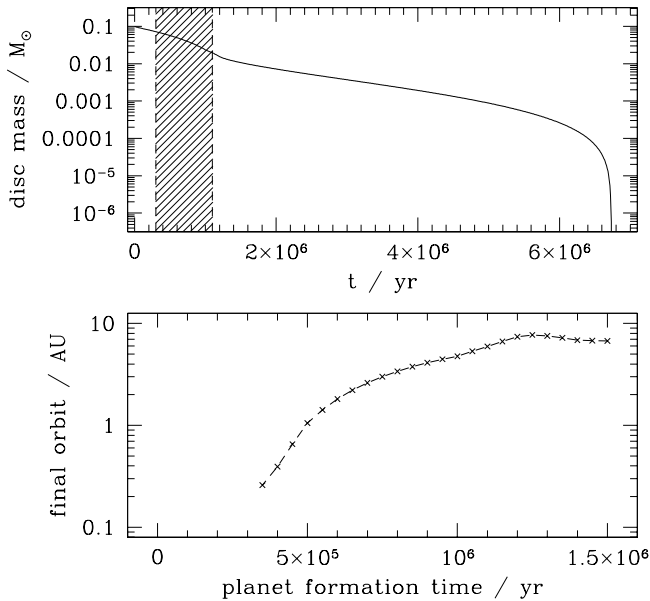,width=3.5truein,height=3.5truein}
\caption{Results of planetary migration in a magnetically layered 
	disc, where we have assumed that the layer is established 
	at $t=1 \ {\rm Myr}$. The upper panel shows the disc 
	mass as a function of time, with the shaded band illustrating 
	the range of planet formation times that result in the 
	planet having a final semi-major axis between 0.1~AU and 
	6~AU. The lower panel shows the final semi-major axis as 
	a function of formation time.}
\label{fig4}	 
\end{figure}

Figure \ref{fig4} shows the results for the runs using the magnetically 
layered disc model. The decline of the disc mass with time shows 
the effect of the switch to a low viscosity state. After the 
initial fully viscous phase, which resembles the standard model, 
there is a long plateau which arises because the low viscosity 
in the inner disc produces a bottleneck to accretion onto the 
star. During this period the disc is slowly destroyed from the 
outside inwards by the disc wind, which 
finally disperses the disc completely after 6~Myr. We note 
that, in a magnetically layered disc model,  
high rates of mass loss from the disc are needed in order to 
disperse the disc within a reasonable time-scale. The 
mass loss rates inferred for discs in parts 
of the Orion nebula are easily large enough 
(Johnstone, Hollenbach \& Bally 1998), 
but it is unknown whether high enough rates are possible 
for isolated discs. 

Figure \ref{fig4} also shows the final location of migrating planets. 
As before, we assume these planets all formed at 5~AU, and 
vary the time of formation in a series of runs. The resulting
$a(t_{\rm form})$ curve flattens for $a \age 1 {\rm AU}$. As 
with the standard disc model, therefore, we predict a 
broad distribution of final orbital radii, with more 
planets (per logarithmic interval in $a$) ending up at 
large radii than at small radii. However, there 
are also important differences. 
Planets must form {\em earlier} (relative to the disc lifetime) 
in the layered disc if they are to migrate successfully to 
small orbital radii. This is because little migration 
can take place once the disc has entered its quiescent, low 
viscosity phase. For a planet to end up inside 1~AU, it 
must have formed at an earlier time when the 
disc was still viscous and able to drive rapid migration.

\section{Comparison with current data}

\begin{figure}
\psfig{figure=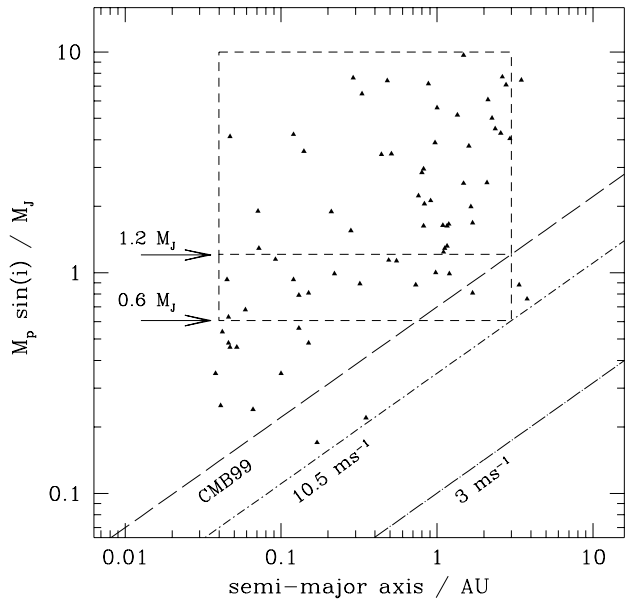,width=3.5truein,height=3.5truein}
\caption{The observed distribution of extrasolar planets in
	the $M_p \sin (i)$ - $a$ plane. The diagonal lines 
	indicate the approximate selection limits of current 
	radial velocity surveys. The upper line, labelled
	CMB99, shows the minimum mass above which planets 
	are detectable at 99\% confidence, according to the 
	analysis of Cumming, Marcy \& Butler (1999). We also 
	plot lines corresponding to radial velocity 
	amplitudes of $3 \ {\rm ms}^{-1}$ and $10.5 \ {\rm ms}^{-1}$.
	The rectangular 
	boxes define subsamples of the data, 
	used for comparison with the predictions of disc migration 
	models.}
\label{fig_data}	 
\end{figure}

Before comparing our results to current data, 
we need to consider possible biases. In an 
idealized radial velocity search for extrasolar planets, two 
important selection effects are expected. First, there is 
almost no sensitivity to planets with orbital periods $P$ 
longer than the duration of the survey (until a whole orbit 
is seen, the planetary signal may be confused with that from 
a more massive, distant companion). Second, a planet with mass 
$M_p$ and orbit inclination $i$ produces a radial velocity signal 
with an amplitude proportional to $M_p \sin (i) a^{-1/2}$. 
A survey which makes a fixed number of observations, with 
some given sensitivity, can therefore detect planets above 
a minimum $M_p \sin(i) \propto a^{1/2}$. Although this is 
a relatively weak radial dependence, it still corresponds to 
an order of magnitude difference in the minimum detectable 
mass over the range of radii probed by current surveys.

Real surveys are less easily characterised. Instruments have 
improved steadily over time, and the intrinsic limits to 
radial velocity measurements vary on a star-by-star basis. 
Nevertheless, the selection effect caused by the 
scaling of the minimum detectable 
mass with $a^{1/2}$ can readily be seen in the data. 
Figure~\ref{fig_data} shows the location of all the 
known extrasolar planets with 
$M_p \sin (i) < 10 \ M_J$, 
in the $M_p \sin (i)$ - $a$ plane. Most of the hot 
Jupiters at $a < 0.1 \ {\rm AU}$, where surveys 
are most sensitive, have masses that are actually 
substantially smaller than that of Jupiter, while 
the typical detected planet with an orbital 
radius greater than 1~AU has a mass of several $M_J$.

\begin{figure}
\psfig{figure=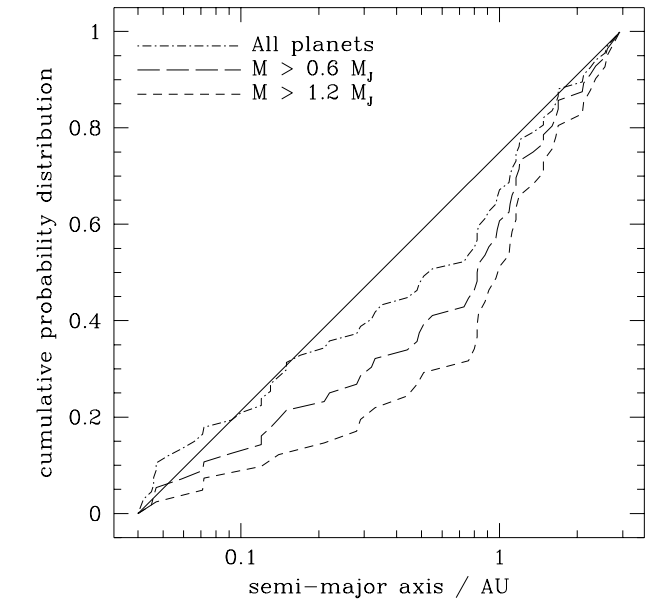,width=3.5truein,height=3.5truein}
\caption{The cumulative probability distribution of extrasolar 
	planets as a function of orbital radius. The dot-dashed curve 
	shows the distribution for all detected planets. This 
	distribution is consistent (KS test probability $P=0.05$) 
	with a uniform distribution in $\log (a)$, shown as the 
	solid line. We also consider subsamples comprising only 
	those planets with masses above  
	$0.6 \ M_J$ (long dashes), or $1.2 \ M_J$ (short dashes). 
	The hypothesis that these subsamples have 
	a distribution that is uniform in $\log (a)$ can 
	be rejected at more than 99.9\% confidence.}
\label{fig_ks}	
\end{figure}

\begin{table}
\begin{tabular}{ccc}
       	& $M_p \sin(i) > 1.2 M_J$ & $M_p \sin(i) > 0.6 M_J$ \\ \hline
       $0.040 - 3.0 \ {\rm AU}$ & $5 \times 10^{-6}$ & $7 \times 10^{-4}$ \\
       $0.065 - 1.9 \ {\rm AU}$ & $10^{-4}$ & $10^{-3}$ \\ \hline
\label{ks_table}
\end{tabular}      
\caption{Kolmogorov-Smirnov test probability that the observed distribution 
	of extrasolar planets is drawn from a distribution uniform in 
	$\log (a)$. The subsamples considered have masses exceeding  
	$0.6 \ M_J$ or $1.2 \ M_J$, and orbital radii within the 
	two quoted ranges.}
\end{table}

A simple count of the total number of detected planets 
as a function of radius leads to the cumulative distribution 
shown in Figure~\ref{fig_ks}. A Kolmogorov-Smirnov test (e.g. 
Press et al. 1989) shows that this is marginally consistent (KS probability 
$P = 0.05$) with a uniform distribution in $\log (a)$. However, 
from Fig.~\ref{fig_data}, it can be seen that this uncorrected 
distribution includes low mass planets, at small orbital radius, 
that would not be detectable beyond 1~AU. The mass function 
of planets has substantial numbers of low mass planets 
(Zucker \& Mazeh 2001; Tabachnik \& Tremaine 2002), so this leads 
to a significant selection effect (Lineweaver \& Grether 2002).

An unbiased estimate of the radial distribution of extrasolar 
planets can be obtained by considering a subsample of 
planets which are sufficiently massive that they could be detected at 
{\em any} radius. A conservative definition of such a subsample 
can be obtained by noting that for the Lick 
planet search, circa 1999, the minimum detectable mass at 99\% 
confidence is quoted 
by Cumming, Marcy \& Butler (1999) as,
\begin{equation}
 M_p \sin (i) \age 0.7 M_J (a/AU)^{1/2}.
\label{eqsensitivity} 
\end{equation} 
Taking this as a guide, a mass cut at $M_{\rm cut} = M_p \sin(i) > 1.2 \ M_J$ 
should yield an approximately complete subsample out to 3~AU.
Since the precision of the observations is likely to have 
improved since 1999, we also consider a subsample with the 
mass cut placed at $0.6 \ M_J$. This corresponds to a 
radial velocity amplitude at 3~AU of $10.5 \ {\rm ms}^{-1}$, which 
is equal to the smallest amplitude signal of any detected 
planet. As one might expect, this practical detection 
threshold lies at $3-4 \sigma$, where $\sigma \approx 3 \ {\rm ms}^{-1}$ 
is the best-case error on a single radial velocity measurement.

As shown in Fig.~\ref{fig_ks} and Table 1, neither of these 
subsamples is consistent with the hypothesis that it is drawn 
from a uniform distribution in $\log (a)$, which can be 
excluded at the 99.9\% confidence level. This conclusion is 
also robust to different choices of 
the minimum and maximum orbital radii of planets in the 
subsample. 
 
To compare with the theoretical distribution derived earlier, we 
henceforth take $M_{\rm cut} = 0.6 \ M_J$. This may somewhat 
{\em underestimate} the number of planets at radii 
of 2-3~AU, but any such bias is smaller than the uncertainties 
introduced by the relatively small sample size. 
We then take the current compilation 
of 76 detected planets (from {\tt exoplanets.org}), 
discard those with masses outside 
the range $0.6 M_J < M_p \sin (i) < 10 M_J$, and bin 
54 survivors into four logarithmic intervals in semi-major 
axis, 0.033~AU -- 0.1~AU, 0.1~AU -- 0.3~AU, 0.3~AU -- 0.9~AU, and 0.9~AU -- 2.7~AU.

\begin{figure}
\psfig{figure=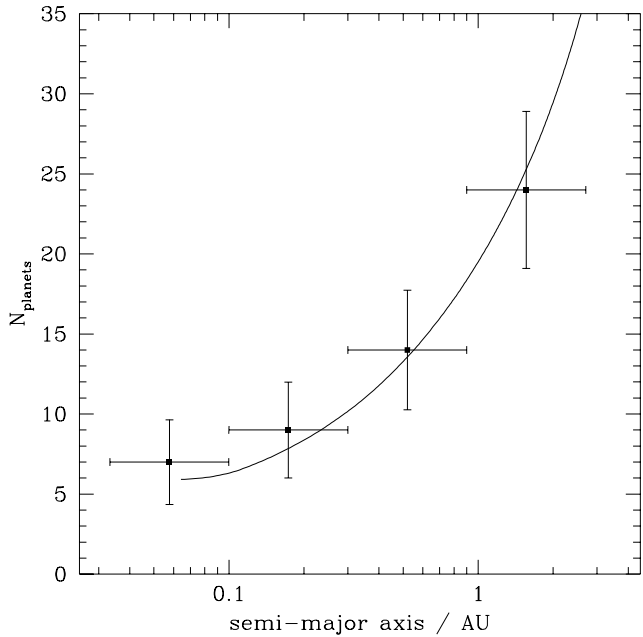,width=3.5truein,height=3.5truein}
\caption{Predicted and observed number of extrasolar planets 
	as a function of radius. The data (points with error 
	bars showing $\sqrt{N}$ errors) show the number of 
	extrasolar planets in four radial bins, using a 
	subsample with a mass cut at $0.6 \ M_J$.	
	The theoretical curve is the prediction 
	for planetary migration within the standard disc 
	model, with the normalisation chosen to give the 
	best fit to the data.}
\label{fig5}	 
\end{figure}

Figure \ref{fig5} shows the comparison between observation 
and theory. The disc inside 0.1~AU may well be strongly 
influenced by the protostellar magnetosphere (Konigl 1991; 
Najita, Edwards, Basri \& Carr 2000), which we have chosen 
not to model. We therefore concentrate attention on the 
outer three radial bins, between 0.1~AU and 2.7~AU.
The number of known extrasolar planets in this 
(approximately) complete subsample rises rapidly with increasing semi-major 
axis. Although the 
error bars on the data are large, excellent agreement 
is obtained with the predictions of the theoretical migration 
model using the standard disc model. This agreement persists 
when different choices of binning and sample selection are made. 
In particular, we can obtain an equally 
good fit (with a different normalization) if we adopt a 
more conservative mass cut at $1.2 \ M_J$.

From the Figure, we also note that in {\em this mass range}, 
the number of hot Jupiters at $a < 0.1 {\rm AU}$ is in fact 
consistent with an extrapolation of the data from larger 
radii. There is no need for an additional stopping 
mechanism to explain the abundance of these 
planets at the smallest radii. There is, however, 
weak evidence for an over-abundance of lower mass planets 
(with $M_p \sin (i) < 0.6 M_J$, not plotted in the Figure) 
in the innermost radial bin. 

\section{Discussion}

\subsection{Frequency of massive planets}

The good agreement between theory and observations seen in 
Figure \ref{fig5} allows us to use the model to estimate 
the fraction of stars targeted by radial velocity surveys 
that possess massive planets, including those with masses 
too low, or semi-major axis too large, to have been detected 
so far. Of course, we cannot extrapolate to masses smaller 
than the $\sim 0.2 M_J \sin(i)$ that is the smallest yet 
detected, as we have no information at all on the mass 
function below this threshold. We can, however, estimate 
how many planets with $0.5 M_J \sin(i)$ 
ought to exist at 1~AU (say), beyond the radius where they 
are currently detectable.

To do so, we note that the radial velocity surveys are most 
complete (i.e. reach down to the smallest masses) at small 
radii. Between 0.1~AU and 0.2~AU, equation (\ref{eqsensitivity}) 
suggests that planets with masses of $0.3 M_J \sin(i)$ and above 
are detectable. 8 planets are known in this radial interval, 
corresponding to approximately 1\% of target stars (since the 
total haul of 76 planets implies that approximately 8\% of stars 
have planets). The range 
of planet formation times which result in the planet becoming 
stranded between 0.1~AU and 0.2~AU is approximately $2.5 \times 10^4 {\rm yr}$, 
which may be compared to the 0.42~Myr window which leaves planets 
between 0.1~AU and 5~AU. Hence, we estimate that the fraction of 
target stars possessing planets 
with masses $0.3 M_{\rm J} < M_{p} \sin (i) < 10 M_{\rm J}$, and radii 
$0.1 {\rm AU} < a < 5 {\rm AU}$, is $\approx$ 15 \%. Roughly half 
of these have already been detected.

More speculatively, we can also try to extrapolate to larger 
radii. If we assume that massive planets at 5-8~AU formed 
near the inner boundary of that region and migrated outwards, 
then a similar analysis to that above suggests that an 
additional 5-10\% of stars could possess such planets. 
This conclusion is more tentative than for inward migration, 
however, because the extent of outward migration varies 
significantly with planet mass (Fig.~\ref{fig2}), and could be 
very different in the magnetically layered disc model.

\subsection{Frequency of massive planet formation}

In our standard model, planets which survive migration 
with final orbital radii between 0.1~AU and 6~AU must 
have formed during a 0.5~Myr allowed window. This window 
lasts for only $\approx$ 20\% of the disc lifetime. Migration 
without a stopping mechanism is thus a moderately inefficient 
process (Trilling et al. 2002) -- several 
planets are likely to have formed for every planet which 
survives today. This inefficiency means that either a large fraction of 
stars formed of the order of one planet, or that a smaller fraction 
of stars formed many planets. Our models do not distinguish 
between these possibilities, though in principle they make 
different predictions. For example, the fraction of stars 
which consume planets (and may show resulting metallicity 
enhancements) is smaller if only a few stars form many 
massive planets. The interpretation of current 
metallicity measurements within such a framework, however,  
remains somewhat controversial (e.g. Gonzalez 1997; Udry, Mayor \& Queloz 2001; 
Murray et al. 2001;  Pinsonneault, DePoy \& Coffee 2001; 
Suchkov \& Schultz 2001). 

\subsection{Number of planets detectable via microlensing}

Detailed monitoring of microlensing events has so far 
failed to find anomalies in the light curves characteristic 
of massive, bound planets (Bennett \& Rhie 1996; Albrow 
et al. 2001; Gaudi et al. 2002). The results 
exclude the possibility that more than around a third 
of lensing stars have Jupiter mass planets at radii 
$1.5 {\rm AU} < a < 4 {\rm AU}$. Our results are consistent 
with this limit. For masses greater than $0.3 M_J$, we 
estimate that 7\% of the stars targeted by radial velocity 
surveys have planets within this range of radii. If this 
is also true of the lensing stars (which are typically 
less massive -- around $0.3 M_\odot$), then it suggests 
that the existing monitoring campaigns are 
within a factor of $\sim 5$ of reaching limits where a 
detection may be expected.

\subsection{Probability of forming a planet which does not migrate}

Unlike in the case of extrasolar planetary systems, the evidence 
for migration in the Solar System is extremely limited. {\em In situ} 
analysis of the Jovian atmosphere shows that it may contain material 
that originated in the outer Solar System (Owen et al. 1999), while 
substantial migration of Uranus and Neptune has been suggested 
(Thommes, Duncan \& Levison 1999). There is little to suggest 
that Jupiter formed a significant distance away from its current location.

Migration models permit both inward and outward 
migration, so forming a planet which does not migrate is perfectly 
possible, though it requires fortuitous timing. If we take 
`no significant migration' to mean that the planet moves a 
distance,
\begin{equation} 
 { {\Delta a} \over a} \leq 0.1,
\end{equation} 
then in our standard disc model the window of formation time 
over which this occurs lasts for $\Delta t = 6.8 \times 10^4 {\rm yr}$.
This may be compared with the 0.5~Myr range of formation times 
that result in the planet having a semi-major axis in the 
range $0.1 {\rm AU} < a < 6 {\rm AU}$. Making the same 
assumption as before -- that the rate of planet formation remains 
constant across this larger interval -- we estimate that no 
significant migration occurs in 
approximately 10-15\% of systems in which a planet survives. The 
no migration outcome which may describe Jupiter in the Solar System 
would then be an unlikely event, but not a rare one (Lineweaver \& 
Grether 2002). Of course forming {\em several} planets, none of 
which migrated significantly, would be much less likely. However, 
as already noted, migration of the other giant planets in the Solar 
System has been seriously considered and does not appear to be 
excluded on observational grounds. 

\subsection{Eccentricity}

Our model for migration ignores the typically substantial 
eccentricities of extrasolar planets.
How serious this omission is depends upon the (unknown) 
mechanism which leads to the eccentricity. If interactions 
between a single planet and the disc are responsible (Artymowicz et al. 
1991; Papaloizou, Nelson \& Masset 2001; Goldreich \& Sari 2002), then a 
description similar to the one we have developed here ought to suffice.
Conversely, eccentricity may arise from planet-planet 
scattering in a multiple planet system (Rasio \& Ford 1996; 
Weidenschilling \& Marzari 1996; Lin \& Ida 1997; 
Papaloizou \& Terquem 2001). This would 
be accompanied by order unity changes in the semi-major 
axis of the surviving planet. The predicted 
distribution of planetary orbital radii, shown in 
Figure~\ref{fig5}, would be changed if all planets (say) 
suffered a shrinkage of their orbits by a factor of two 
subsequent to migration. We have no additional need to invoke 
such a process to explain the orbital radii of the planets, but 
more work in this area is warranted.

\section{Conclusions}

We have shown that the radial distribution of massive extrasolar 
planets, beyond 0.1~AU, is consistent with the planets forming 
at 5~AU, before migrating inwards through a viscous protoplanetary 
disc. This means that although theory suggests that  
massive planets could form closer to the star (Papaloizou \& 
Terquem 1999; Bodenheimer, Hubickyj \& Lissauer 2000; 
Sasselov \& Lecar 2000), there is no {\em observational} 
requirement for planets to form at $a < 3 {\rm AU}$.   
The predicted distribution is a robust feature of 
disc models which include dispersal of the gas via a disc 
wind, so the level of comparison is currently limited by 
the small numbers of detected planets.
The continuation of existing radial velocity surveys, along with 
forthcoming astrometric searches (e.g. Lattanzi et al. 2000), 
will allow for more stringent 
tests, provided that the selection effects of the surveys are 
well understood.

The good agreement between observations and theory allows 
us to use the model to estimate the fraction of stars 
that possess massive planets, including those currently 
undetected on account of their low mass or large semi-major 
axis. Limiting ourselves to planets with masses 
$0.3 M_{\rm J} < M_{p} \sin (i) < 10 M_{\rm J}$, and radii 
$0.1 {\rm AU} < a < 5 {\rm AU}$, we estimate that around 
15\% of stars in the current target sample of extrasolar 
planet searches possess planets. In some of our models 
a significant number of planets migrate {\em outwards}, 
and these could form a sizeable additional population. 
These numbers are comparable to those found by  
Trilling, Lunine \& Benz (2002), and imply that even in 
existing samples, most of the 
massive planets remain to be discovered.

Finally, we have used the results to quantify the fraction 
of systems in which the most massive planet does not suffer 
significant radial migration. This allows us to address 
the concern that migration theories could be 
inconsistent with the detailed knowledge of our own Solar 
System, in which Jupiter may have formed close to its 
current location. We find that a no-migration outcome 
requires fortuitous timing of the epoch of planet formation, 
but not extraordinary luck. Of the order of 10-15\% of planetary 
systems are expected to have a massive planet which has migrated 
by less than 10\% in orbital radius. Having a most massive 
planet at the orbital radius of Jupiter is thus expected 
to occur in around 1-2\% of solar-type stars. The orbital radius of 
Jupiter, at least, is consistent with our 
understanding of extrasolar planets, though the 
near-circular orbit may yet prove to be unusual.

\section*{Acknowledgments}

PJA thanks the Institute of Astronomy for their hospitality 
during the course of this work. JEP is grateful to STScI for 
continuing support under their Visitors Program. SL acknowledges 
support from NASA grants NAG5-4310 and NAG5-10732. We thank 
the anonymous referee for a detailed and helpful review.

\end{document}